\documentclass[12pt,fleqn]{article}
\usepackage{amsmath,amssymb}
\usepackage{bm}
\usepackage[dvips]{graphicx}

\begin{document}
\begin{center}

{\bf \Large Quantum  Gravity without General Relativity }

\vspace{1cm}

Takehisa Fujita\footnote{e-mail:
fffujita@phys.cst.nihon-u.ac.jp} 

Department of Physics, Faculty of Science and Technology, 

Nihon University, Tokyo, Japan

\vspace{2cm}

{\Large Abstract}

\end{center}

The quantum field theory of gravitation is constructed in terms of Lagrangian 
density of Dirac fields which couple to the electromagnetic field $A_\mu$ as well as the 
gravitational field $\cal G$. The gravity appears in the mass term as 
$ m(1+g {\cal G}) \bar{\psi}\psi $ with the coupling constant of $g$. In addition to 
the gravitational force between fermions, the electromagnetic field $A_\mu$ interacts 
with the gravity as the fourth order effects and its strength amounts to $\alpha$ 
times the gravitational force.  Therefore, the interaction of photon with gravity 
is not originated from Einstein's general relativity which is entirely dependent on 
the unphysical assumption of the principle of equivalence. Further, we present a 
renormalization scheme for the gravity and show that the graviton stays massless.

\newpage

\section{ Introduction $-$ Problems of General Relativity}
The motion of the earth is governed by the gravitational force between 
the earth and the sun, and the Newton equation is written as
$$ m \ddot{\bm r} =-G_0mM{\bm{r}\over r^3} \eqno{(1.1)} $$
where $G_0$, $m$ and $M$ denote the gravitational constant, the mass of the earth and 
the mass of the sun, respectively. This is the classical mechanics which works quite well. 
Further, Einstein generalizes the Newton equation to the relativistic 
equation of motion which can be valid even for the curved space \cite{ein,wein}. 
However, this was achieved before the discovery of quantum mechanics, and therefore 
it is natural that the general relativity cannot be quantized properly. Indeed, 
the quantization of the general relativity has intrinsic problems which are related 
to the invariance of the general coordinate transformation. On the other hand, 
the first quantization is only possible for the Cartesian coordinates \cite{fujita}. 
This indicates that any attempt to quantize the general relativity is not a proper 
starting point, but rather one should try to make a field theory simply to 
include the gravity. This is closely connected to the understanding of the first 
quantization ($ [x_i, p_j] = i\hbar \delta_{ij}$), and since this quantization 
procedure is not a fundamental principle, we should try to make a field theory 
which includes the gravitational interaction \cite{fujita,fks2}. 

Before constructing a theory that can describe the field equation under the gravity, 
we discuss the fundamental problems in the theory of general relativity. Basically, 
there are two serious problems in the general relativity, the lack of field equation 
under the gravity and the assumption of the principle of equivalence. 

\subsection{Field Equation of Gravity}
When one wishes to write the Dirac equation for a particle under 
the gravitational interaction, then one faces to the difficulty. 
Since the Dirac equation for a hydrogen-like atom can be written as 
$$ \left(-i \bm{\nabla}\cdot \bm{\alpha}+m\beta -{Ze^2\over r} \right) \Psi 
=E\Psi \eqno{(1.2a)}  $$
one may write the Dirac equation for the gravitational potential 
$ V(r)= -{G_0mM\over r}$  as
$$ \left(-i \bm{\nabla}\cdot \bm{\alpha}+m\beta -{G_0mM\over r} \right) \Psi 
=E\Psi. \eqno{(1.2b)}  $$
But there is no foundation for this equation. At least, one cannot write the Lagrangian 
density which can describe the Dirac equation for the gravitational interaction. This  
is clear since one does not know whether the interaction can be put into the zero-th 
component of a vector type or a simple scalar type in the Dirac equation. 
That is, it may be of the following type
$$ \left[-i \bm{\nabla}\cdot \bm{\alpha}+ \left(m -{G_0mM\over r}\right) 
\beta \right] \Psi =E\Psi. \eqno{(1.2c)}  $$
This is a well known problem, but it is rarely discussed, and 
people seem to be reluctant to treating this problem up on the table. 

\subsection{Principle of Equivalence}
The theory of general relativity is entirely based on the principle of equivalence. 
Namely, Einstein started from the assumption that physics of the two systems 
(a system under the uniform external gravity and a system that moves with a constant 
acceleration) must be the same. This looks plausible from the experience on the earth. 
However, one can easily convince oneself that the system that moves with a constant 
acceleration cannot be defined properly since there is no such an isolated system 
in a physical world. The basic problem is that the assumption of the principle 
of equivalence is concerned with the two systems  which specify space and time, 
not just the numbers in connection with the acceleration of a particle. Note that 
the acceleration of a particle is indeed connected to the gravitational acceleration, 
$ \ddot z = -g $, but this is, of course, just the Newton equation. 
Therefore, the principle of 
equivalence inevitably leads Einstein to the space deformation. It is clear that 
physics must be the same between two inertia systems, and any assumption which contradicts 
this basic principle cannot be justified at all. 

Besides, this problem can be viewed differently in terms of Lagrangian. 
For the system under the uniform external gravity, one can write the corresponding 
Lagrangian. On the other hand, there is no way to construct any Lagrangian for the system 
that moves with a constant acceleration. One can define a Lagrangian for a particle 
that moves with a constant acceleration, but one cannot write the system 
(or space and time) that moves with a constant acceleration. 
Therefore, it is very hard to accept the assumption of the principle of equivalence 
even with the most modest physical intuition. 

\subsection{General Relativity}
Einstein generalized the classical mechanics to the relativistic equation of 
motion where he started from the principle of equivalence. Therefore, he had to introduce 
the new concept that space may not be uniform, and the general relativity is the equation 
for the metric tensor $g_{\mu \nu}$. However, this picture is still based on the particle 
mechanics which is governed by the equations for coordinates of a point particle. 
This is, of course, natural for Einstein since quantum mechanics was not discovered 
at that time. 
Since quantum mechanics is a field theory, though with the non-relativistic kinematics, 
it is essentially different from Newton's classical mechanics but is rather similar 
to the Maxwell equations. Newton equation can certainly describe the dynamics of particles 
for the certain region of kinematics such as the motion of the earth around the sun. 
However, it is a useless theory for the description of electron motion in atoms. 
One should give up the idea of particle picture and should accept the concept of 
field theory. At the time of invention of the general relativity, Einstein knew quite 
well the Maxwell equations which are indeed field theory equations. However, the Maxwell 
equations are not realized as the basis equations for quantum mechanics \cite{fujita}.

\vspace{0.5cm}
\section{Lagrangian Density for Gravity}
It is by now clear that one should start from constructing the quantum mechanics 
of the gravitation. In other words, one should find the Dirac 
equation for electron when it moves in the gravitational potential. 
In this paper, we present a model Lagrangian density which can describe electrons 
interacting with the electromagnetic field $A_\mu$ as well as the gravitational 
field $\cal G$. 

\subsection{Lagrangian Density for QED}
We first write the well established Lagrangian density for electrons 
interacting 
with the electromagnetic field  $A_\mu$ 
$$  {\cal L}_{el} =  i\bar \psi  \gamma^{\mu}{\partial}_{\mu}\psi -e\bar \psi\gamma^{\mu}
 A_{\mu}\psi -m \bar \psi \psi   -{1\over 4} F_{\mu\nu} F^{\mu\nu}   \eqno{ (2.1)} $$
where
$$   F_{\mu\nu}= \partial_{\mu} A_{\nu} - \partial_{\nu} A_{\mu} . $$
This Lagrangian density of QED is best studied and is most reliable in many respects. 
In particular, the renormalization scheme of QED is theoretically well understood and 
is experimentally well examined, and there is no problem at all in the perturbative 
treatment of QED. All the physical observables can be described in terms of the free 
Fock space terminology after the renormalization, and therefore one can compare 
any prediction of the physical quantities with experiment. However, it should be 
noted that QED is the only field theory model in four dimensions which works perfectly 
well without any conceptual difficulties.

\subsection{Lagrangian Density for QED plus Gravity}
Now, we propose to write the Lagrangian density for electrons interacting 
with the electromagnetic field as well as the gravitational field $\cal G$ 
$$  {\cal L} =  i\bar \psi  \gamma^{\mu}{\partial}_{\mu}\psi -e\bar \psi\gamma^{\mu}
 A_{\mu}\psi -m(1+g{\cal G}) \bar \psi \psi 
  -{1\over 4} F_{\mu\nu} F^{\mu\nu}  
+ {1\over 2}\partial_\mu {\cal G} \ \partial^\mu {\cal G} \eqno{ (2.2)} $$
where the gravitational field ${\cal G}$ is assumed to be a massless scalar field. 
It is easy to prove that the new Lagrangian density is invariant under the local 
gauge transformation 
$$ A_\mu \rightarrow A_\mu +\partial_\mu \chi, \ \ \ \psi \rightarrow 
e^{-ie\chi} \psi . \eqno{ (2.3)} $$
This is, of course, quite important since the introduction of the gravitational 
field does not change the most important local symmetry. 

\subsection{Dirac Equation with Gravitational Interactions }
Now, one can easily obtain the Dirac equation for electrons from the new Lagrangian 
density
$$    i \gamma^{\mu}{\partial}_{\mu}\psi -e \gamma^{\mu} A_{\mu}\psi 
-m(1+g{\cal G})  \psi =0 . \eqno{ (2.4)}  $$
Also, one can write the equation of motion of gravitational field 
$$  \partial_\mu\partial^\mu  {\cal G} =- mg \bar \psi \psi . \eqno{ (2.5)} $$
The symmetry property of the new Lagrangian density can be easily examined, and one can 
confirm that it has a right symmetry property under the time reversal transformation, 
parity transformation and the charge conjugation \cite{fujita}. 

\subsection{Total Hamiltonian for QED plus Gravity}

The Hamiltonian can be constructed from the Lagrangian density in eq.(2.2)
$$ H=\int \left\{ \bar{\psi}  \left(-i \bm{\gamma}\cdot \bm{\nabla} 
+m(1+g{\cal G})  \right)  \psi
-e \bm{j}\cdot \bm{A}  \right\}d^3r  
 +{e^2\over 8\pi} \int {j_0(\bm{r}')j_0(\bm{r})d^3rd^3r'
\over{|\bm{r}'-\bm{r}| }} $$
$$ + {1\over 2}\int  \left(\dot{\bm{A}}^2+
(\bm{\nabla \times \bm{A}})^2   \right) d^3r  
 +{1\over 2}\int   \left(\dot{{\cal G}}^2+(\bm{\nabla } {\cal G})^2 \right)d^3r
 \eqno{(2.6)}  $$
where $j_\mu$ is defined as $ j_\mu =\bar{\psi}\gamma_\mu \psi $. 
In this expression of the Hamiltonian, the gravitational energy is still written 
without making use of the equation of motion. In the next section, we will treat 
the gravitational energy and rewrite it into an expression which should enable us 
to easily understood the structure of gravitational force between fermions. 

\vspace{0.5cm}
\section{Static-dominance Ansatz for Gravity}
In eq.(2.2), the gravitational field ${\cal G}$ is introduced as a {\it real scalar} 
field, and therefore it cannot be a physical observable 
as a classical field \cite{kkof}. 
In this case, since the real part of the right hand side in eq.(2.5) should be mostly time 
independent, it may be reasonable to assume that the gravitational field ${\cal G}$  
can be written as the sum of the static and time-dependent terms and that the static part 
should carry the information of diagonal term in the external source term. 
Thus, the gravitational field ${\cal G}$ is assumed to be written as
$$ {\cal G}={\cal G}_0(\bm{r} )+\bar{\cal G}(x) \eqno{(3.1)} $$
where ${\cal G}_0(\bm{r} )$ does not depend on time. This ansatz is only a sufficient 
condition, and its validity cannot be verified mathematically, but it can be 
examined experimentally. 

The equations of motion 
for ${\cal G}_0(\bm{r} )$ and $\bar{\cal G}(x)$  become
$$ \bm{ \nabla}^2  {\cal G}_0 = mg \rho_g \eqno{ (3.2)} $$
$$  \partial_\mu\partial^\mu  \bar{\cal G}(x) = - mg 
\{ (\bar \psi \psi)_{\rm [non-diagonal]}
+ (\bar{\psi} \psi)_{\rm [diagonal \ rest]} \} \eqno{ (3.3)} $$
where $\rho_g$ is defined as
$$ \rho_g \equiv (\bar{\psi} \psi)_{\rm [diagonal]}  \eqno{ (3.4)} $$
where $(\bar{\psi} \psi)_{\rm [diagonal]}$ denotes the diagonal part of the $\bar{\psi} 
\psi$, that is, the terms proportional to $ [ a^{\dagger (s)}_{\bm{k}}a^{(s)}_{\bm{k}'}- 
b^{\dagger (s)}_{\bm{k}}b^{(s)}_{\bm{k}'} ]$ of the fermion operators which will be 
defined in eq.(4.2). Further, $(\bar \psi \psi)_{\rm [non-diagonal]}$ term is 
a non-diagonal part which is connected to the creation and annihilation of fermion pairs, 
that is,  $ [ a^{\dagger (s)}_{\bm{k}}b^{\dagger (s)}_{-\bm{k}'}+ 
b^{ (s)}_{-\bm{k}'}a^{(s)}_{\bm{k}} ]$ of the fermion operators.  In addition, 
the term $(\bar{\psi} \psi)_{\rm [diagonal \ rest]}$ denotes  time dependent 
parts of the diagonal term in the fermion density, and this may also have some effects 
when the gravity is quantized. 
In this case, we can solve eq.(3.2) exactly and find a solution 
$$ {\cal G}_0(\bm{r} )=-{mg\over 4\pi} \int {\rho_g(\bm{r}')
\over{|\bm{r}'-\bm{r}| }} d^3r' \eqno{ (3.5)} $$
which is a special solution that satisfies eq.(2.5), but not the general solution. 
Clearly as long as the solution can satisfy the equation of motion of eq.(2.5), 
it is physically sufficient. The solution of eq.(3.5) is quite important for 
the gravitational interaction since this is practically a dominant gravitational force 
in nature. 

Here, we assume that the diagonal term of $(\bar \psi \psi)_{\rm [diagonal]}$ 
is mostly time independent, and in this case, the static gravitational energy 
which we call $H_G^{S}$ can be written as
$$ H_G^{S}= mg \int \rho_g {\cal G}_0 d^3r +{1\over 2}\int (\bm{\nabla } {\cal G}_0)^2 
 d^3r  $$
$$ =-{m^2 G_0\over 2} \int {\rho_g(\bm{r}')\rho_g(\bm{r})
\over{|\bm{r}'-\bm{r}| }} d^3rd^3r' \eqno{(3.6a)}  $$
where the gravitational constant $G_0$ is related to the coupling constant $g$ as
$$ G_0={g^2\over 4\pi}  . \eqno{(3.7)}  $$
This static gravitational energy can be written in the momentum representation as
$$ H_G^{S}= -{m^2 G_0\over 4\pi^2} \sum_{\bm{p},\bm{p}' }
\int {\bar{u}(\bm{p}+\bm{q})u(\bm{p}) \bar{u}(\bm{p}'-\bm{q})u(\bm{p}') 
\over{q^2 }} d^3q.   \eqno{(3.6b)}  $$
Eq.(3.6) is just the gravitational interaction energy for the matter fields, 
and one sees that the gravitational interaction between electrons 
is always attractive. This is clear since the gravitational field is assumed 
to be a massless scalar. It may also be important to note that the  $H_G^{S}$ of eq.(3.6) 
is obtained without making use of the perturbation theory, and it is indeed exact, 
apart from the static ansatz of the field ${\cal G}_0(\bm{r} )$.

\vspace{0.5cm}

\section{Quantization of Gravitational Field}
In quantum field theory, we should quantize fields. For fermion fields, we should quantize 
the Dirac field by the anti-commutation relations of fermion operators. This is required 
from the experiment in terms of the Pauli principle, that is, a fermion can occupy 
only one quantum state. In order to accommodate this experimental fact, we should always 
quantize the fermion fields with the anti-commutation relations. On the other hand, 
for gauge fields, we must quantize 
the vector field in terms of the commutation relation which is also required 
from the experimental observation that one photon is 
emitted by the transition between $2p-$state and $1s-$state in hydrogen atoms. 
That is, a photon is created from the vacuum of the electromagnetic field, and 
therefore the field quantization is an absolutely necessary procedure. 
However, it is not very clear whether the gravitational field ${\cal G}$ 
should be quantized according to the bosonic commutation relation or not. In fact, 
there must be two choices concerning the quantization of the gravitational field ${\cal G}$. 

\subsection{No Quantization of Gravitational Field $\bar{\cal G}$}
As the first choice, we may take a standpoint that the gravitational field ${\cal G}$ 
should not be quantized since there is no requirement from experiments. 
In this sense, there is no definite reason that we have to quantize the scalar 
field and therefore the gravitational field ${\cal G}$ should remain to be a classical 
field. In this case, we do not have to worry about the renormalization of the graviton 
propagator, and we obtain the gravitational interaction between fermions as we saw 
it in eq.(3.6) which is always attractive, and this is consistent with the experimental 
requirement. 

\subsection{Quantization Procedure}
Now, we take the second choice and should 
quantize the gravitational field $\bar{\cal G}$. This can be 
done just in the same way as usual scalar fields
$$ \bar{\cal G} (x) =\sum_{\bm{ k}} {1\over{\sqrt{2V\omega_k}}}\left[
d_ke^{-i\omega_k t+i \bm{ k}\cdot \bm{ r} }+
d_k^\dagger e^{i\omega_k t-i \bm{ k}\cdot \bm{ r} } \right] \eqno{(4.1)} $$
where $\omega_k =|\bm{k}| $. The annihilation and creation operators 
$d_k $ and $d_k^\dagger $ are assumed to satisfy the following commutation 
relations
$$ [d_{\bm{k}} , d_{\bm{ k}'}^\dagger ]= \delta_{\bm{k},\bm{k}' } \eqno{(4.2)} $$
and all other commutation relations should vanish.  Since the graviton can couple to 
the time dependent external field which is connected to the creation or annihilation 
of the fermion pairs, the graviton propagator should be affected from the vacuum 
polarization of fermions. Therefore, we should carry out the renormalization procedure of 
the graviton propagator such that it can stay massless.  We will discuss 
the renormalization procedure in the later chapter. 

\subsection{Graviton}
Once the gravitational field ${\cal G}$ is quantized, then the graviton should appear. 
From eq.(4.1), one can see that the graviton can indeed propagate as a free massless 
particle after it is quantized, and this situation is just the same as the gauge 
field case in QED, namely, photon after the quantization becomes a physical observable. 
However, it should be noted that the gauge field has a special feature in the sense that 
the classical gauge field ($\bm{A}$) is gauge dependent and therefore it is 
not a physical observable. After the gauge fixing, the gauge field can be quantized 
since one can uniquely determine the gauge field from the equation of motion, and 
therefore its quantization is possible. 

On the other hand, the gravitational field is assumed to be a real scalar field, and 
therefore it cannot be a physical observable as a classical field \cite{kkof}. 
Only after 
the quantization, it becomes a physical observable as a graviton, and this can be seen 
from eq.(4.1) since the creation of the graviton should be made through the second term of 
eq.(4.1). In this case, the graviton field is a complex field which is an eigenstate of 
the momentum and thus it is a free graviton state, which can propagate as a free particle. 

\vspace{0.5cm}

\section{Interaction of Photon with Gravity}
From the Lagrangian density of eq.(2.2), one sees that photon should interact 
with the gravity in the fourth  order Feynman diagrams as shown in Fig. 1. 
The interaction Hamiltonian $H_I$ can be written as
$$ H_I=\int \left( mg  {\cal G}  \bar{\psi}  \psi
-e \bar{\psi} \bm{\gamma} \psi \cdot \bm{A}  \right) d^3r \eqno{(5.1)} $$
where the fermion field $\psi$ is quantized 
in the normal way
$$ \psi (\bm{r},t) = \sum_{\bm{p},s} {1\over{\sqrt{L^3}}}
\left( a_{{\bm{p}}}^{(s)} u^{(s)}_{{\bm{p}}}
e^{i\bm{p}\cdot \bm{r}-i E_{\bm{p}} t }  
 +{b}^{\dagger (s)}_{{\bm{p}}} v^{ (s)}_{{\bm{p}}}
e^{-i\bm{p}\cdot \bm{r} +iE_{\bm{p}} t}  \right)  \eqno{(5.2)}  $$
where $u^{(s)}_{{\bm{p}}}$ and $v^{(s)}_{{\bm{p}}}$
denote the spinor part of the plane wave solutions of the free Dirac equation. 
$a^{(s)}_{{\bm{p}}}$ and $b^{(s)}_{{\bm{p}}}$ are
annihilation operators for particle and anti-particle states, and
they should satisfy the following anti-commutation relations,
$$ \{ a^{(s)}_{{\bm{p}}},
{a^\dagger}_{{\bm{p}}'}^{(s')} \} =\delta_{s, s'}
\delta_{{{\bm{p}}}, {{\bm{p}}'} }, \ \ 
\{ b^{(s)}_{{\bm{p}}},
{b^\dagger}_{{\bm{p}}'}^{(s')} \} =\delta_{s, s'}
\delta_{{{\bm{p}}}, {{\bm{p}}'} }  \eqno{(5.3)}  $$
and all other anticommutation relations should vanish. 
The gauge field $\bm{A}$  can be quantized in the same way
$$ \bm{A}(x)=\sum_{\bm{k}} \sum_{\lambda =1}^2{1\over{\sqrt{2V\omega_{\bm{k}}}}}
\bm{\epsilon}^\lambda(\bm{k}) \left[ c_{\bm{k},\lambda} e^{-ikx} +
  c^{\dagger}_{\bm{k},\lambda} e^{ikx} \right]  \eqno{(5.4)} $$
where $ \omega_{\bm{k}}=|\bm{k}| $. 
The polarization vector $ \bm{\epsilon}^\lambda(\bm{k}) $ should satisfy the following
relations
$$ \bm{\epsilon}^\lambda(\bm{k})\cdot \bm{k} =0, \ \ \ \
\bm{\epsilon}^\lambda(\bm{k}) \cdot \bm{\epsilon}^{\lambda'}(\bm{k}) =\delta_{\lambda,
\lambda'} . \eqno{(5.5)}  $$
The annihilation and creation operators  
$c_{\bm{k},\lambda}$, \  $c_{\bm{k},\lambda}^\dagger $ should satisfy the following 
commutation relations
$$ [ c_{\bm{k},\lambda}, \  c_{\bm{k}',\lambda'}^\dagger  ] = \delta_{ \bm{k}, \bm{k}' }
\delta_{\lambda, \lambda'} \eqno{(5.6)}  $$
and all other commutation relations should vanish.

\begin{figure}\centering
\unitlength 0.1in
\begin{picture}( 35.0500, 26.9000)( -1.3500,-34.2500)
%
\special{pn 8}%
\special{pa 284 344}%
\special{pa 254 352}%
\special{pa 226 340}%
\special{pa 216 312}%
\special{pa 232 284}%
\special{pa 244 274}%
\special{sp}%
%
\special{pn 8}%
\special{pa 284 344}%
\special{pa 314 336}%
\special{pa 344 346}%
\special{pa 352 374}%
\special{pa 338 402}%
\special{pa 326 412}%
\special{sp}%
%
\special{pn 8}%
\special{pa 366 480}%
\special{pa 336 488}%
\special{pa 308 478}%
\special{pa 298 450}%
\special{pa 314 422}%
\special{pa 326 412}%
\special{sp}%
%
\special{pn 8}%
\special{pa 366 480}%
\special{pa 396 472}%
\special{pa 426 484}%
\special{pa 436 512}%
\special{pa 420 540}%
\special{pa 408 550}%
\special{sp}%
%
\special{pn 8}%
\special{pa 450 618}%
\special{pa 420 626}%
\special{pa 390 616}%
\special{pa 380 586}%
\special{pa 396 560}%
\special{pa 408 550}%
\special{sp}%
%
\special{pn 8}%
\special{pa 450 618}%
\special{pa 480 610}%
\special{pa 508 620}%
\special{pa 518 650}%
\special{pa 502 676}%
\special{pa 490 686}%
\special{sp}%
%
\special{pn 8}%
\special{pa 532 754}%
\special{pa 502 764}%
\special{pa 472 752}%
\special{pa 464 724}%
\special{pa 478 696}%
\special{pa 490 686}%
\special{sp}%
%
\special{pn 8}%
\special{pa 532 754}%
\special{pa 562 746}%
\special{pa 590 758}%
\special{pa 600 788}%
\special{pa 584 814}%
\special{pa 572 824}%
\special{sp}%
%
\special{pn 8}%
\special{pa 614 892}%
\special{pa 584 900}%
\special{pa 554 890}%
\special{pa 546 862}%
\special{pa 560 834}%
\special{pa 572 824}%
\special{sp}%
%
\special{pn 8}%
\special{pa 614 892}%
\special{pa 644 884}%
\special{pa 672 894}%
\special{pa 682 924}%
\special{pa 668 950}%
\special{pa 654 960}%
\special{sp}%
%
\special{pn 8}%
\special{pa 288 2588}%
\special{pa 258 2580}%
\special{pa 230 2590}%
\special{pa 220 2618}%
\special{pa 236 2646}%
\special{pa 248 2656}%
\special{sp}%
%
\special{pn 8}%
\special{pa 288 2588}%
\special{pa 318 2596}%
\special{pa 348 2586}%
\special{pa 356 2556}%
\special{pa 342 2528}%
\special{pa 330 2520}%
\special{sp}%
%
\special{pn 8}%
\special{pa 370 2450}%
\special{pa 340 2442}%
\special{pa 312 2454}%
\special{pa 302 2482}%
\special{pa 318 2510}%
\special{pa 330 2520}%
\special{sp}%
%
\special{pn 8}%
\special{pa 370 2450}%
\special{pa 400 2458}%
\special{pa 430 2448}%
\special{pa 440 2418}%
\special{pa 424 2392}%
\special{pa 412 2382}%
\special{sp}%
%
\special{pn 8}%
\special{pa 454 2314}%
\special{pa 424 2306}%
\special{pa 394 2316}%
\special{pa 384 2344}%
\special{pa 400 2372}%
\special{pa 412 2382}%
\special{sp}%
%
\special{pn 8}%
\special{pa 454 2314}%
\special{pa 484 2322}%
\special{pa 512 2310}%
\special{pa 522 2282}%
\special{pa 506 2254}%
\special{pa 494 2244}%
\special{sp}%
%
\special{pn 8}%
\special{pa 536 2176}%
\special{pa 506 2168}%
\special{pa 476 2178}%
\special{pa 468 2208}%
\special{pa 482 2236}%
\special{pa 494 2244}%
\special{sp}%
%
\special{pn 8}%
\special{pa 536 2176}%
\special{pa 566 2184}%
\special{pa 594 2174}%
\special{pa 604 2144}%
\special{pa 588 2116}%
\special{pa 576 2108}%
\special{sp}%
%
\special{pn 8}%
\special{pa 618 2038}%
\special{pa 588 2030}%
\special{pa 558 2042}%
\special{pa 550 2070}%
\special{pa 564 2098}%
\special{pa 576 2108}%
\special{sp}%
%
\special{pn 8}%
\special{pa 618 2038}%
\special{pa 648 2046}%
\special{pa 676 2036}%
\special{pa 686 2008}%
\special{pa 672 1980}%
\special{pa 658 1970}%
\special{sp}%
%
\special{pn 8}%
\special{ar 372 1464 600 600  5.2528085 6.2831853}%
\special{ar 372 1464 600 600  0.0000000 1.0303768}%
%
\special{pn 8}%
\special{ar 990 1464 600 600  2.1112158 4.1719695}%
%
\special{pn 8}%
\special{pa 970 1460}%
\special{pa 1570 1460}%
\special{da 0.070}%
\special{sh 1}%
\special{pa 1570 1460}%
\special{pa 1504 1440}%
\special{pa 1518 1460}%
\special{pa 1504 1480}%
\special{pa 1570 1460}%
\special{fp}%
%
\special{pn 8}%
\special{pa 1570 1460}%
\special{pa 2170 1460}%
\special{da 0.070}%
%
\special{pn 8}%
\special{pa 2170 1460}%
\special{pa 2570 860}%
\special{fp}%
\special{sh 1}%
\special{pa 2570 860}%
\special{pa 2516 904}%
\special{pa 2540 904}%
\special{pa 2550 928}%
\special{pa 2570 860}%
\special{fp}%
%
\special{pn 8}%
\special{pa 2570 860}%
\special{pa 2970 260}%
\special{fp}%
%
\special{pn 8}%
\special{pa 2970 2660}%
\special{pa 2570 2060}%
\special{fp}%
\special{sh 1}%
\special{pa 2570 2060}%
\special{pa 2590 2128}%
\special{pa 2600 2104}%
\special{pa 2624 2104}%
\special{pa 2570 2060}%
\special{fp}%
%
\special{pn 8}%
\special{pa 2570 2060}%
\special{pa 2170 1460}%
\special{fp}%
%
\special{pn 20}%
\special{sh 1}%
\special{ar 970 1460 10 10 0  6.28318530717959E+0000}%
\special{sh 1}%
\special{ar 2170 1460 10 10 0  6.28318530717959E+0000}%
\special{sh 1}%
\special{ar 680 950 10 10 0  6.28318530717959E+0000}%
\special{sh 1}%
\special{ar 680 1980 10 10 0  6.28318530717959E+0000}%
%
\special{pn 8}%
\special{pa 390 1490}%
\special{pa 390 1430}%
\special{fp}%
\special{sh 1}%
\special{pa 390 1430}%
\special{pa 370 1498}%
\special{pa 390 1484}%
\special{pa 410 1498}%
\special{pa 390 1430}%
\special{fp}%
%
\special{pn 8}%
\special{pa 850 1100}%
\special{pa 890 1160}%
\special{fp}%
\special{sh 1}%
\special{pa 890 1160}%
\special{pa 870 1094}%
\special{pa 860 1116}%
\special{pa 836 1116}%
\special{pa 890 1160}%
\special{fp}%
%
\special{pn 8}%
\special{pa 900 1740}%
\special{pa 860 1800}%
\special{fp}%
\special{sh 1}%
\special{pa 860 1800}%
\special{pa 914 1756}%
\special{pa 890 1756}%
\special{pa 880 1734}%
\special{pa 860 1800}%
\special{fp}%
\put(15.7000,-16.1000){\makebox(0,0){$\bm{q}$}}%
\put(29.7000,-28.1000){\makebox(0,0){$\bm{p}$}}%
\put(2.7000,-28.1000){\makebox(0,0){$\bm{k}$}}%
\put(2.7000,-1.2000){\makebox(0,0){$\bm{k}'$}}%
\put(29.7000,-1.2000){\makebox(0,0){$\bm{p}'$}}%
\put(16.4000,-33.0000){\makebox(0,0){Fig. 1: The fourth order Feynman diagram}}%
\end{picture}%

\end{figure}

The calculation of the S-matrix can be carried out in a straightforward way 
\cite{bd,q2,nishi}, and we can write 
$$ S=(ie)^2 \epsilon_\mu^\lambda(k) \epsilon_\nu^{\lambda'}(k')  
\left({mm'g^2\over q^2}\right) \bar{u}(p') u(p) $$ 
$$ \times \int { d^4a\over{(2\pi)^4}} {\rm Tr} 
\left[ \gamma_\mu {i\over{a \llap/ -m+i\epsilon}}
\gamma_\nu {i\over{b \llap/ -m+i\epsilon}} {i\over{c \llap/ -m+i\epsilon}} \right] 
 \eqno{(5.7)}   $$
where $k$ and $k'$ denote the four momenta of the initial and final photons while  
$p$ and $p'$ denote the four momenta of the initial and final fermions, respectively. 
$m$ and $m'$ denote the mass of the fermion for the vacuum polarization and 
the mass of the external fermion. 
$a$, $b$, $c$ and $q$ can be written in terms of $k$ and $p$ as
$$ q=p'-p, \ \ \ \ k=a-b, \ \ \ \ k'=a-c, \ \ \ \ q=k-k'. $$
Therefore, the S-matrix can be written as
$$ S=ie^2 mm' g^2 \epsilon_\mu^\lambda(k) \epsilon_\nu^{\lambda'}(k') 
{1\over q^2}\bar{u}(p') u(p)  
\int { d^4a\over{(2\pi)^4}} {1\over a^2-m^2} {1\over (a-k)^2-m^2} 
{1\over (a-k')^2-m^2} $$
$$ \times {\rm Tr} \left[ \gamma_\mu (a \llap/ +m) 
\gamma_\nu  ((a\llap/- k\llap/+m) ((a\llap/-k\llap/'+m)\right]. \eqno{(5.8)}  $$
Since the term proportional to $q$ does not contribute to the interaction, 
we can safely approximate in the evaluation of the trace and the $a$ integration as
$$ k' \approx k . $$
Now, we define the trace part as
$$ N_{\mu \nu} ={\rm Tr} \left[ \gamma_\mu (a \llap/ +m) 
\gamma_\nu  ((a\llap/- k\llap/+m) ((a\llap/-k\llap/'+m)\right] \eqno{(5.9)}  $$
which can be evaluated as
$$ N_{\mu \nu} =4m[ (k^2-a^2+m^2)g_{\mu \nu}+4a_\mu a_\nu -2a_\mu k_\nu 
-2a_\nu k_\mu ] . \eqno{(5.10)}  $$
Defining the integral by
$$ I_{\mu \nu} \equiv  \int { d^4a\over{(2\pi)^4}} {N_{\mu \nu}
\over{( a^2-m^2) \left[ (a-k)^2-m^2 \right] 
\left[ (a-k')^2-m^2 \right] } } \eqno{(5.11)} $$
we can rewrite it using Feynman integral
$$ I_{\mu \nu} = 2\int { d^4a\over{(2\pi)^4}} \int_0^1 zdz 
{N_{\mu \nu}\over{ [ (a-kz)^2-m^2+z(1-z)k^2]^3 }}. \eqno{(5.12)} $$ 
Therefore, introducing the variable $ w=a-kz$ we obtain the S-matrix as
$$ S=8ie^2 m^2m' g^2 \epsilon_\mu^\lambda(k) \epsilon_\nu^{\lambda'}(k') 
{1\over q^2}\bar{u}(p') u(p) \times  $$ 
$$ \int { d^4w\over{(2\pi)^4}} 
\left[ { (-w^2g_{\mu \nu}+4w_\mu w_\nu)\over{[ w^2-m^2+z(1-z)k^2]^3 }} 
 +  { \left\{ m^2 +k^2(1-z^2)\right\} g_{\mu \nu}+4k_\mu k_\nu z(1-z)
\over{[ w^2-m^2+z(1-z)k^2]^3 }} \right]. \eqno{(5.13)} $$
The first part of the integration can be carried out in a straightforward way 
using the dimensional regularization, and we find
$$ \int { d^4w\over{(2\pi)^4}} 
 { (-w^2g_{\mu \nu}+4w_\mu w_\nu)\over{[ w^2-m^2+z(1-z)k^2]^3 }} =   
 {i\pi^2\Gamma(0)g_{\mu \nu}\over 2\Gamma (3)}(4-4)=0. $$
Thus, the two divergent parts just cancel with each other, and the cancellation 
here is not due to the regularization as employed  
in the vacuum polarization in QED, but it is a kinematical and thus rigorous result. 
The finite part can be easily evaluated \cite{nishi}, and therefore we obtain the S-matrix 
as
$$ S={e^2\over 8\pi} m^2m' g^2 (\epsilon^\lambda\epsilon^{\lambda'}) 
{1\over q^2}\bar{u}(p') u(p) \eqno{(5.14)}  $$ 
where we made use of the relation $k^2=0$ for free photon at the end of the calculation. 

\vspace{0.5cm}
\section{Renormalization Scheme for Gravity}
At the present stage, it is difficult to judge whether we should quantize 
the gravitational field or not. At least, there is no experiment which shows any 
necessity of the quantization of the gravity. Nevertheless, it should be worth checking 
whether the gravitational interaction with fermions can be renormalizable or not. 
We know that the interaction of the gravity with fermions is extremely small, but 
we need to examine whether the graviton can stay massless or not within 
the perturbation scheme. 

Here, we present a renormalization scheme for the scalar field theory which couples 
to fermion fields. The renormalization scheme for scalar fields is formulated just in the same way as the QED scheme since QED is most successful. 

\subsection{Vacuum Polarization of Gravity}
First, we write the vacuum polarization for QED with the dimensional 
regularization, and the divergent contributions to the self-energy of photon can be 
described in terms of the vacuum polarization $\Pi^{\mu \nu}_{QED}(k)$ as
$$ \Pi^{\mu \nu}_{QED}(k)=i\lambda^{4-D}e^2\int {d^Dp\over(2\pi)^D}
{\rm Tr} \left[ \gamma^\mu {1\over p \llap/-m  } \gamma^\nu
{1\over p \llap/-k \llap/-m  }\right] $$
$$ ={e^2\over 6\pi^2\epsilon}(k^\mu k^\nu-g^{\mu \nu}k^2) + 
{\rm finite \ terms} \eqno{(6.1)}  $$
where $D$ is taken to be $D=4-\epsilon$. 
It is interesting to note that the apparent quadratic divergence disappears 
due to the gauge invariant dimensional regularization when evaluating the momentum 
integrations. This is important since, if there were any quadratic divergence terms 
present, then it would have caused serious troubles for the mass terms which cannot keep 
the gauge invariance in QED. The fact that the quadratic divergence terms can be 
erased by the proper dimensional regularization in QED is indeed related to the success 
of QED renormalization scheme. 

On the other hand, the vacuum 
polarization for the gravity becomes 
$$ \Pi(k)=i\lambda^{4-D}m^2g^2\int {d^Dp\over(2\pi)^D}
{\rm Tr} \left[  {1\over p \llap/-m  } {1\over p \llap/-k \llap/-m  }\right] $$
$$ = {m^2g^2\over 12\pi^2}\left\{ 3\Gamma (-1+{\epsilon \over 2} ) 
 \left(m^2-{1\over 6}k^2\right)
+\Gamma ({\epsilon } )k^2 \right\} \eqno{(6.2)}  $$
This can be rewritten as 
$$ \Pi(k)=-{1\over 2}C_1 k^2 -{1\over 2}C_2 m^2 \eqno{(6.3)}  $$
where 
$$ C_1 =-{m^2g^2\over 2\pi^2}\left( {1\over \epsilon}-{\gamma\over 2}+{1\over 6} 
\right) \eqno{(6.4a)}  $$
$$ C_2 ={m^2g^2\over 2\pi^2}\left( {2\over \epsilon}-{\gamma}+1  
\right). \eqno{(6.4b)}  $$
As can be seen, the second term in eq.(6.3) should correspond to the quadratic 
divergence term, and this time it cannot be erased by the dimensional regularization. 
However, this term can be safely eliminated by the counter term. Therefore, we add 
the following Lagrangian density as the mass counter-terms to the original Lagrangian 
density 
$$ \delta {\cal L} ={1\over 2}C_1 \partial_\mu {\cal G} \ \partial^\mu {\cal G}  
- {1\over 2}\delta M  {\cal G}^2 \eqno{(6.5)}  $$
where the constant $\delta M $ is defined as 
$$ \delta M \equiv C_2 m^2. \eqno{(6.6)} $$
Therefore, the total Lagrangian density of the gravity $  {\cal L}_G $ becomes 
$$  {\cal L}_G ={1\over 2}(1+C_1 )\partial_\mu {\cal G} \ \partial^\mu {\cal G}  
- {1\over 2}\delta M  {\cal G}^2 
={1\over 2}\partial_\mu {\cal G}_r \ \partial^\mu {\cal G}_r  
- {1\over 2}\delta M  {\cal G}_r^2  \eqno{(6.7)}  $$
where ${\cal G}_r$ is the renormalized gravity field. This shows that the mass counter 
term cannot be renormalized into the wave function ${\cal G}$. However, the mass counter 
term in eq.(6.7) has a proper symmetry property of the gravity Lagrangian density, 
in contrast to the QED case where the mass term violates the gauge invariance. Therefore, 
the introduction of the mass counter term in the scalar field theory does not break 
the renormalization scheme of the present formulation. 

\subsection{Fermion Self Energy from Gravity}
The fermion self energy term in QED is calculated to be
$$ \Sigma_{QED}(p)=-ie^2 \int{d^4k\over (2\pi)^4}
\gamma_\mu { 1\over p \llap/-k \llap/-m }\gamma^\mu { 1\over k^2 }
={e^2\over 8\pi^2\epsilon}(-p \llap/ +4m)+
\textrm{finite terms} . \eqno{(6.8)} $$
In the same way, we can calculate the fermion self-energy due to the gravity
$$ \Sigma_G(p)=-im^2g^2\lambda^{4-D}\int {d^Dk\over(2\pi)^D}
 { 1\over p \llap/-k \llap/-m } { 1\over k^2 } ={m^2g^2\over 8\pi^2\epsilon}(-p \llap/ +4m)
+ \textrm{finite terms}   \eqno{(6.9)}  $$
which is just the same as the QED case, apart from the factor in front. 
Therefore, the renormalization procedure can be carried out just in the same 
way as the QED case since the total fermion self energy term within the present model 
becomes
$$ \Sigma(p)={1\over 8\pi^2\epsilon}(e^2+m^2g^2)(-p \llap/ +4m)
+ \textrm{finite terms}  . \eqno{(6.10)}  $$

\subsection{Vertex Correction from Gravity}
Concerning the vertex corrections which arise from the gravitational interaction and 
electromagnetic interaction with fermions, it may well be that the vertex corrections 
do not become physically very important. It is obviously too small to measure 
any effects of the higher order terms from the gravity and electromagnetic interactions. 
However, we should examine the renormalizability of the vertex corrections 
and can show that they are indeed well renormalized into the coupling constant. 
The vertex corrections from the electromagnetic interaction and the gravity can be 
evaluated as 
$$ \Lambda_{QED}(k,q)=i\lambda^{4-D}mge^2\int {d^Dp\over(2\pi)^D}
{\rm Tr} \left[ \gamma_\mu  {1\over (k \llap/-p \llap/-m)
(k \llap/-p \llap/-q \llap/-m) p^2 } \gamma^\mu \right] 
 \eqno{(6.11a)}  $$
$$ \Lambda_G(k,q)=i\lambda^{4-D}m^3g^3\int {d^Dp\over(2\pi)^D}
{\rm Tr} \left[   {1\over (k \llap/-p \llap/-m)(k \llap/-p \llap/-q \llap/-m) p^2 }\right]. 
 \eqno{(6.11b)}  $$
We can  easily calculate the integrations and obtain the total vertex corrections 
for the zero momentum case of $q=0$ as
$$ \Lambda(k,0)=\Lambda_{QED}(k,0)+ \Lambda_G(k,0)
={mg\over \pi^2\epsilon}(e^2+m^2g^2) + {\rm finite \ \ terms} 
\eqno{(6.12}  $$
which is logarithmic divergence and is indeed renormalizable just in the same way 
as the QED case. 

\subsection{Renormalization Procedure}
Since the infinite contributions to the fermion self-energy 
and to the vertex corrections in the second order diagrams are just the same as the QED 
case, one can carry out the renormalization procedure just in the same 
way as the QED case. There is only one difference between QED and the gravity cases, 
that is, the treatment of the quadratic divergence in the vacuum polarization. 
In the QED case, the quadratic divergence terms should be eliminated by the dimensional 
regularization since the mass term violates the gauge invariance and thus one cannot 
consider the mass counter term in the QED Lagrangian density. On the other hand, 
in the gravity case, 
the quadratic divergence terms in the vacuum polarization can be canceled out by a mass 
counter term since the gravity is not the gauge field theory, and thus, there is 
no problem to introduce the mass counter term in the Lagrangian density. 
Further, the graviton is never bound and always 
in the free state, and therefore, the mass counter term in the gravity cancels 
the quadratic divergence contribution in a rigorous wayDIn this way, we can achieve 
a successful renormalization scheme for the gravity, even though we do not know any 
occasions in which the higher order contributions may become physically important.

\vspace{0.5cm}
\section{Gravitational Interaction of Photon with Matter }

From eq.(5.14), one finds that the gravitational potential $V(r)$ for photon with 
matter field can be written as
$$ V(r)=-{ G_0\alpha m^2_t M \over 2}{1\over r} \eqno{(7.1)}  $$ 
where $m_t$ and $M$ denote the sum of all the fermion masses and the mass of matter field, 
respectively. $\alpha$ denotes the fine structure constant $\alpha ={1\over 137}$. 
In this case, the equation of motion for photon $\bm{A}_\lambda $ under 
the gravitational field becomes
$$ \left( {\partial^2\over{\partial t^2}}-\bm{\nabla}^2 -{ G_0\alpha m^2_t M\over 2}
{1\over r} \right)\bm{A}_\lambda =0 . \eqno{(7.2)}  $$
Assuming the time dependence of the photon field  $\bm{A}_\lambda $ as
$$  \bm{A}_\lambda =\bm{\epsilon}_\lambda e^{-i\omega t} A_0(\bm{r}) \eqno{(7.3)} $$
we obtain 
$$ \left( -\bm{\nabla}^2 -{ G_0\alpha m^2_t M \over 2}{1\over r} 
 \right)A_0(\bm{r})=\omega^2 A_0(\bm{r}). \eqno{(7.4)}  $$
This equation shows that there is no bound state for photon even for the strong 
coupling limit of $G_0 \rightarrow \infty$.  

\vspace{0.5cm} 
\section{ Conclusions}

We have presented a new scheme of treating the gravitational interactions 
between fermions in terms of the Lagrangian density. The gravitational 
interaction appears always as the mass term and induces always the attractive 
force between fermions. In addition, there is an interaction between photon 
and the gravity as the fourth order Feynman diagrams. The behavior of photon 
under the gravitational field may have some similarity with the result 
of the general relativity, but the solution of eq.(7.4) is still to be studied 
in detail. 

Also, we have presented a renormalization procedure which is essentially the same 
as the QED renormalization scheme. There is one important difference between 
the QED and the gravity cases, that is, the treatment of the quadratic 
divergence in the vacuum polarization. In QED, one has to eliminate the quadratic 
divergence terms by the regularization so as to keep the gauge invariance of 
the Lagrangian density. On the other hand, in the gravity case, the quadratic 
divergence terms can be canceled out 
by the mass counter term since it does not contradict with any important symmetry of 
the Lagrangian density. Therefore, the renormalization scheme of the gravity interaction 
is well justified, and thus the propagator of the gravity stays massless.  
Clearly, this is the most important point in the whole renormalization procedure. 

In this paper, we have not decided whether the gravitational field should be 
quantized or not since there is no definite requirement from experiment 
for the quantization. At the present stage, both of the evaluation of 
the gravitational interactions with fermions should be equally reasonable. 
However, for the quantized theory of gravitational field, one may ask as to whether 
there is any method to observe a graviton or not. The graviton should be created 
through the fermion pair annihilation. Since this graviton can propagate as a free 
graviton like a photon, one may certainly have some chance to observe it through 
the creation of the fermion pair. But this probability must be extremely small since 
the coupling constant is very small, and there is no enhancement in this process 
unless a strong gravitational field like a neutron star may rapidly change as a function 
of time.

\ \ 

\vspace{0.1cm}
\section*{ACKNOWLEDGEMENTS}

We thank Prof. K. Nishijima for encouragements and helpful comments. 
In particular, the photon-gravity vertex part and the renormalization 
procedure for the scalar field are clarified a great deal 
through discussions. 

\vspace{1.0cm}


\end{document}